\documentstyle[12pt]{article}
\newlength{\abstractwidth}
\begin{document}
\thispagestyle{empty}
\pagestyle{plain}
\renewcommand{\thefootnote}{\fnsymbol{footnote}}
\renewcommand{\thanks}[1]{\footnote{#1}} 
\newcommand{\starttext}{
\setcounter{footnote}{0}
\renewcommand{\thefootnote}{\arabic{footnote}}}

\begin{titlepage}
\bigskip
\hskip 3.7in\vbox{\baselineskip12pt\hbox{FERMILAB-PUB-95/099-T}
\hbox{NSF-ITP-94-37}\hbox{hep-th/9505054}}
\bigskip\bigskip\bigskip\bigskip

\centerline{\Large\bf Maximally Supersymmetric String Theories in $D$$<$$10$}

\bigskip\bigskip

\centerline{\bf Shyamoli Chaudhuri\thanks{sc@itp.ucsb.edu}}
\medskip
\centerline{Institute for Theoretical Physics}
\centerline{University of California}
\centerline{Santa Barbara, CA\ \ 93106-4030}

\bigskip\bigskip

\centerline{\bf George Hockney
and Joseph D. Lykken\thanks{hockney@fnalv.fnal.gov; lykken@fnalv.fnal.gov}}
\medskip
\centerline{Fermi National Accelerator Laboratory}
\centerline{P.O. Box 500}
\centerline{Batavia, IL\ \ 60510}

\bigskip\bigskip

\begin{abstract}\baselineskip=16pt
The existence of maximally supersymmetric solutions to heterotic string
theory that are not toroidal compactifications of the ten-dimensional
superstring is established. We construct an exact fermionic realization of
an N=1 supersymmetric string theory in $D$$=$$8$ with non-simply-laced gauge
group $Sp(20)$. Toroidal compactification to six and four dimensions gives
maximally extended supersymmetric theories with reduced rank $(4,12)$ and
$(6,14)$ respectively.
\end{abstract}
\end{titlepage}
\starttext
\baselineskip=18pt
\setcounter{footnote}{0}

Finiteness is a robust property of the perturbative amplitudes of the known
superstring theories.
N=4 supersymmetric Yang-Mills theory is known to be
finite in four dimensions \cite{seiberg}, and there is growing evidence that
the theory exhibits an extension of
Olive-Montonen strong-weak coupling duality
known as S-duality\cite{gno} \cite{gira}.
A generalization of the Olive-Montonen
duality of N=4 theories has also been identified in N=1 supersymmetric
Yang-Mills theory \cite{sei}. In string theory,
conjectures for S-duality have mostly been explored in the context of
toroidal compactifications of the ten-dimensional heterotic string to
spacetime dimensions $D$$<$$10$ \cite{sen}.

It would be helpful to have insight into the generic
moduli space, and the generic duality group, of such {\it maximally
supersymmetric} string theories. We will therefore consider the possibility
of exact solutions to string theory beyond those obtained by dimensional
reduction from a ten-dimensional superstring.
These solutions are exact in the sigma model ($\alpha^{\prime}$) expansion
but are perturbative in the string coupling constant.
To be specific, we will construct solutions to heterotic string
theory, i.e., with $(N_R,N_L)$$=$$(2,0)$ superconformal invariance
on the world-sheet. Our construction
is, however, quite general and the conclusions can be adapted to
solutions of any closed string theory in any spacetime dimension.

Toroidal compactification of the ten-dimensional N=1 heterotic string to
six (four) dimensions  results in a low-energy
effective N=2 (N=4) supergravity coupled to $20$ ($22$) abelian vector
multiplets, giving a total of $24$ ($28$) abelian vector gauge fields
with gauge group $(U(1))^{24}$ ($(U(1))^{28}$), respectively.
Four (six) of these abelian multiplets are contained within the
$N$$=$$2$ ($N$$=$$4$) supergravity multiplets. At enhanced symmetry
points in the moduli space the abelian group $(U(1))^{20}$
$((U(1))^{22})$ is enlarged to a simply-laced group of
rank $20$ ($22$). The low energy field theory limit of such a solution
has maximally extended spacetime supersymmetry. Since all of the
elementary scalars appear in the adjoint representation
of the gauge group, symmetry breaking via the Higgs mechanism is
adequate in describing the moduli space of vacua with a
{\it fixed} number of abelian multiplets.

In this work we show that there exist maximally supersymmetric
vacua with four, six, and eight-dimensional Lorentz invariance that
are not obtained by toroidal compactification of a ten-dimensional heterotic
string. The total number of abelian vector multiplets in the
four-dimensional theory can be reduced
to just {\it six}, namely, those contained within the N=4 supersymmetry
algebra. This is consistent with known theorems on the world-sheet
realizations of extended spacetime supersymmetry in string theory\cite{bd}.
In the world-sheet description of an N=4 supersymmetric solution
of the heterotic string in four dimensions, the internal
right moving superconformal field theory of central charge $c_R$$=$$9$
is required to be composed of nine free bosons.
A reduction of the rank of the low-energy gauge group in an N=4 solution
implies that the internal left-moving conformal field theory of central charge
$c_L$$=$$22$ is not entirely composed of free bosons. This is unlike the
4D toroidal compactifications described by Narain \cite{narain} where
{\it both} right
and left moving conformal field theories are free boson theories.

As a consequence, it will also be possible to realize
non-simply-laced gauge symmetry consistent with the maximally
extended supersymmetries. We will construct such solutions using
real fermionization \cite{klst}. Exact solutions to heterotic
string theory obtained in this construction are examples
of rational $(2,0)$ superconformal field theories, where the underlying
chiral algebras also have a world-sheet fermionic realization. In order
to have an unambiguous identification of the vertex operator algebra
in the fermionic construction, it is essential to have explicit knowledge
of the correlators of the real fermion conformal field theories \cite{cchl}.

Eliminating longitudinal and time-like modes, the
number of transverse degrees of freedom describing a vacuum with D-dimensional
Lorentz invariance is $(c_R,c_L)$$=$$({3\over2}\cdot (D-2),D-2)$$+$
$(c^{int}_R, c^{int}_L)$. In this class of exact solutions, the internal
degrees of freedom have an {\it equivalent} world-sheet fermionic realization
with $(3\cdot (10-D), 2\cdot (26-D))$ Majorana-Weyl fermions. The
world-sheet fermionic realization is convenient, both as an explicit
calculational tool and because it allows us to construct,
consistent with finiteness and anomaly cancellation, an exact solution to
string theory which embeds {\it a specified low-energy matter content}.

We restrict ourselves to fermionic realizations where the
world-sheet fermions are Majorana-Weyl, with periodic or anti-periodic
boundary conditions only.
All of the right-moving world-sheet fermions will be paired into
Weyl fermions, or equivalently free bosons, as required by the extended
spacetime supersymmetry. A free boson conformal field theory implies,
with no loss of generality, the existence of an abelian current in the
right-moving superconformal field theory. In maximally supersymmetric
solutions the allowed right-moving chiral algebras are, therefore,
restricted to level one simply-laced affine Lie
algebras \cite{narain} \cite{bd}. This follows
from the fact that for a Lie algebra with roots of equal length, the
central charge of the level one realization also equals the rank of the
algebra, i.e., the number of abelian currents.

A free fermionic realization with $n$ Weyl (complex) fermions exists
for any of the following affine Lie algebras: $SO(2n)$,
$U(n)$, and $E_8$ (for $n$$=$$8$), in addition to the abelian
algebra $(U(1))^n$.
In toroidal compactifications that have an equivalent free fermionic
realization these
properties also extend to the allowed left-moving chiral algebras and,
hence, to the observed non-abelian gauge symmetry in these solutions.

Incorporating {\it real} fermion world-sheet fields in the left-moving
internal conformal field theory will enable us to construct maximally
supersymmetric solutions that embed non-simply-laced gauge symmetry,
i.e., gauge groups with roots of unequal length. Such solutions
necessarily lie in a moduli space where the gauge group has
rank $<$ $28$. This is evident from the formula for the central charge
of an affine Lie algebra:
\begin{equation}
c ~~=~~ {{k ~ Dim(G)}\over{k + {\tilde h}}}
\label{central charge}
\end{equation}
\noindent where the dual Coxeter number, $\tilde h $, of the
non-simply-laced algebras, $SO(2n+1)$, $Sp(2n)$, $G_2$ and $F_4$ are,
respectively,
$2n-1$, $n+1$, $4$, and $9$. Note that the dimension of the dual
algebras $SO(2n+1)$ and $Sp(2n)$ are identical, given by $Dim(G)$$=$$n(2n+1)$.
However, unlike the simply-laced algebras, the central charge does
not equal the rank of the group even at level $k$$=$$1$, and does not,
in fact, coincide for the algebra and its dual.
Real fermion realizations exist for all of the non-simply-laced
affine algebras.
Extending a world-sheet fermionic realization of the
generators of the affine algebra to a $(2,0)$ superconformal field
theory that is an exact solution to heterotic string
theory, however, requires consistency with modular invariance of the one-loop
vacuum amplitude and with world-sheet supersymmetry \cite{cchl}. These
conditions can
be quite restrictive and, in fact, preclude N=1 supersymmetric solutions in
ten spacetime dimensions with non-simply-laced gauge symmetry.

Now consider the possibility of non-simply-laced gauge symmetry in
$D$$<$$10$. For example, an affine realization of the rank ten algebra
$Sp(20)$ at level one requires central charge $c$$=$${{35}\over{2}}$.
Appending a single real fermion with $c$$=$${1\over 2}$ gives
$c$$=$$18$, making this a plausible candidate for the gauge group of an
N=1 spacetime supersymmetric solution in $D$$=$$8$. It is not
difficult to verify the existence of such a solution using its
fermionic realization.

We will adopt the notation of \cite{klst}\cite{cchl}. The tree level spectrum
is described by the one-loop vacuum amplitude, which sums over sectors
labelled by the associated spin structure of the world-sheet fermions.
The N=1 spacetime supersymmetry charges are embedded in the spin-structure
of eight right-moving Majorana-Weyl fermions, which we will label
$\psi_{\mu}$, $\mu$$=$$1$, $\cdots$, $6$, $\psi_7$, and $\psi_{10}$.
The spin-${3\over 2}$ generator of the $(1,0)$ world-sheet
supersymmetry is the operator
\begin{equation}
T_F({\bar{z}}) =
i \sum_{\mu = 1}^{6} \psi_{\mu} \partial_{\bar z} X^{\mu}
{}~~+~~ i \sum_{k=2,3} \psi_{3k+1} \psi_{3k+2} \psi_{3k+3}
\label{supercurrent}
\end{equation}
\noindent The first six right-movers therefore carry a (transverse) spacetime
index. In sectors contributing spacetime bosonic and fermionic
components of an N=1 supermultiplet, these eight fermions are, respectively,
Neveu-Schwarz and Ramond. In particular the {\it untwisted} sector,
${\cal U}$, in which
all of the world-sheet fermions are Neveu-Schwarz, contributes the
bosonic components of the N=1 supergravity multiplet in eight
dimensions. It also contributes two massless abelian multiplets, each
associated with an internal right-moving Weyl fermion:
$\psi_8+i \psi_{11}$, and $\psi_9+i \psi_{12}$.
Thus the full gauge group of this model will be
$Sp(20)$$\times$$(U(1))^2$. Note that the Ramond vacuum of the right-moving
fermions $\psi_7$,
$\psi_{10}$ is constrained by modular invariance
to be aligned with that of the first six right-movers. Thus,
there cannot exist modular invariant solutions to heterotic string
theory with extended spacetime supersymmetry in $D$$=$$8$,
as expected from the viewpoint of the low-energy effective Lagrangian.

The remaining massless spectrum is arranged into $D$$=$$8$
N=1 Yang-Mills supermultiplets, each
containing 6 spacetime vector components, 8 spinor components,
and 2 scalar components \cite{salam}.
The sector-wise decomposition of the 210 states in the adjoint
representation of $Sp(20)$ is most easily described
by the regular embedding:
\begin{equation}
Sp(20) ~\supset ~ (SO(5))^5 ~\supset ~(SO(4))^5 ~ \sim ~ (SU(2))^{10}
\label{embedding}
\end{equation}
\noindent The untwisted sector, ${\cal U}$, contributes states
corresponding to all 30 long roots, and a subset (20)
of the short roots of $Sp(20)$. These states transform,
respectively, in the adjoint (10 copies of a $\bf 3$) and the
spinor (10 copies of a doublet) representation of
its $(SU(2))^{10}$ sub-group.
The states are identified by fermionic charge: the
roots and weights of the rank ten sub-group are embedded in
the fermionic charge of ten {\it Weyl} fermions. In the
fermionic construction these are obtained by pairing
20 Majorana-Weyl left-movers,
$\psi_{2l+1}(z) + i \psi_{2l+2}(z)$$=$$\lambda_{l}(z)$, $l$$=$$0, \cdots
9$.

The remaining 16 left-moving Majorana-Weyl fermions are {\it real}
fermions. The vertex operator construction for an $SO(2n+1)$ algebra
requires a single real fermion, in addition to $n$ Weyl fermions.
The long-root lattice of $SO(2n+1)$ coincides with the root-lattice of
$SO(2n)$, ${\bf \Lambda_L}(B_n)$$=$$D_n$. Thus the $n$$\cdot$$(2n$$-$$1)$
Majorana-Weyl fermion bilinears are the currents corresponding to
long roots, while those corresponding to the short roots are the $2n$
bilinears containing the single real fermion.
In this example, of the ${{20 \cdot 19}\over{2}}$
Neveu-Schwarz fermion bilinear currents contributed by the untwisted
sector only ${{ 5 \cdot 5 \cdot 4}\over{2}}$ remain after GSO projection
from four {\it twisted} sectors,
${\cal T}_1$ $\cdots$ ${\cal T}_4$, in
which some of the fermions are Ramond. The untwisted sector
therefore contributes a total of 400 states: the eight bosonic components
of an N=1 supermultiplet transforming in the adjoint representation of
the non-simply-laced group $(SO(5))^5$.

Extension of this vertex operator construction to a symplectic
current algebra requires conformal dimension $(h_R,h_L)$$=$$(0,1)$ operators
corresponding to the additional short roots. These are contributed by the
twisted sectors. The currents are composite operators
constructed out of sixteen {\it twist} fields, i.e., dimension
$(0,{{1}\over{16}})$ operators
in the Majorana-Weyl fermion field theory.

The twisted sectors, ${\cal T}_i$, were chosen
so as to generate the necessary projection on the untwisted sector.
They will simultaenously determine the internal right-moving chiral
algebra: in this solution, the four internal right-moving
fermions, $\psi_8$, $\psi_9$, $\psi_{11}$, and $\psi_{12}$, are
either all Neveu-Schwarz, or all Ramond, in every sector of the Hilbert
space. Thus the underlying right-moving chiral algebra is $SO(4)$.
Possible twists are, of course, subject to constraints from modular
invariance and world-sheet supersymmetry.
Given a set of valid ${\cal T}_i$, modular invariance
of the one-loop vacuum amplitude automatically generates additional
twisted sectors in the Hilbert space. Thus, in this example, the
${\cal T}_i + {\cal T}_j$, $i$$\ne$$j$, also contribute massless
states in the spectrum. Each of the ten twisted sectors
contributes 128 states:
8 bosonic components of an N=1 supermultiplet transforming
in the 16 dimensional spinor representation of an $(SU(2))^4$ sub-group.
$Sp(20)$ has ten distinct $(SU(2))^4$ sub-groups, each
corresponding to a different twisted sector. Combining the
400 untwisted sector states with these 1280 states gives all
$8$$\cdot$$210$ bosonic components of an N=1
supermultiplet transforming in the adjoint representation
of $Sp(20)$.

It is straightforward to construct the twisted sector vertex operator
corresponding to a given weight. We will use the
bosonic realization for the corresponding free field
vertex operator. A state transforming as a spinor
weight, $\alpha$, of $(SU(2))^4$ corresponds to a dimension
$(0,{{1}\over{2}})$ operator, $j_{\rm free}(z)$,
obtained by bosonization:
\begin{equation}
\lambda_l^{\dagger} \lambda_l ~\leftrightarrow ~ \partial \phi_l
{}~~~~~j_{\rm free}(z) ~=~ {\hat C} (\alpha) ~ e^{i \alpha \cdot \phi}
\end{equation}
\noindent where $\alpha \cdot \alpha $$=$$1$,
$l$$=$$0$, $\cdots$ $9$, and the ${\hat C}(\alpha)$ are suitable
cocycle operators. This free field vertex operator must
be dressed by four pseudo-Weyl fermion spin fields,
$\sigma_l^{\pm}$, $ l=1, ~\cdots ~ 4$, so as to give a current. These
spin fields are identified
by pseudo-complexifying, i.e., pairing, the real fermions in a
twisted sector \cite{cchl}. Thus:
\begin{equation}
J_{ijkl}(z) =j_{free} (z)
{}~ \left( \sigma_i^+\sigma_j^+\sigma_k^+\sigma_l^+
{}~+ ~ \sigma_i^-\sigma_j^-\sigma_k^-\sigma_l^- \right)
\label{current}
\end{equation}
\noindent where $ i \ne j \ne k \ne l $, giving a dimension $(0,1)$
twisted sector current. Verification of the vertex operator algebra
for $Sp(20)$ is now straightforward.

This completes the discussion of the massless spectrum of the
N=1 supersymmetric $Sp(20)$ heterotic string in eight dimensions.
Anomaly cancellation is particularly simple in this theory:
there is no gravitational anomaly in $D$$=$$8$ dimensions
\cite{agwit}, the right-moving $U(1)$'s are non-anomalous,
and $Sp(20)$ is an anomaly-free gauge group.
Compactification on a torus will give anomaly-free theories
with maximally extended spacetime supersymmetry in lower dimensions.
Quite generally, compactification on an $(n,n)$ dimensional $D_n$
lattice has an equivalent fermionic realization in terms of $(n,n)$
Weyl fermions. Recall that the fermionic description of the
$Sp(20)$ string theory contained an extra left-moving real fermion.
Appending this real fermion to the Weyl fermion realization of the
$SO(2n)$ current algebra extends it to a realization of $SO(2n+1)$.
It is straightforward to verify, as we have done,
the existence of an N=2 $Sp(20)$$\times$$SO(5)$ solution in six
dimensions and an N=4 $Sp(20)$$\times$$SO(9)$ solution in four
dimensions with fermionic realizations.
Thus toroidal compactification of the eight-dimensional N=1
$Sp(20)$ heterotic string gives an N=4 theory in four dimensions with
only {\it twenty} abelian vector multiplets, or rank $(6,14)$, at
generic points in the moduli space.
The target space duality group clearly has an
$O(4,4;Z)$$\backslash$$O(4,4)$$/$$(O(4)\times O(4))$
subgroup corresponding to the moduli space of the torus, but it
will be extended by the background modes of the D=8
supergravity-Yang-Mills theory \cite{narain}.

The moduli spaces of six-dimensional solutions are of particular
interest in exploring string-string duality. The conjectured S-duality of the
heterotic string compactified on a six-dimensional torus has been shown to
follow as a consequence of target space T-duality of the type
IIA string theory: compactified on the K3 surface, this string theory
is dual to the heterotic string compactified on a four-dimensional
torus \cite{town} \cite{witten} \cite{harstrom}. As described above,
toroidal compactification of the ten-dimensional $E_8$$\times$$E_8$ string
and the eight-dimensional $Sp(20)$ string gives rank 24 and rank 16 moduli
spaces, respectively. Twisting the $Sp(20)$$\times$$SO(5)$ solution gives
a solution with exceptional gauge symmetry: $F_4$$\times$$F_4$$\times$$Sp(8)$.
It
is likely that these solutions belong to the same moduli space. We
have also constructed a new family of N=2 solutions with 12 abelian multiplets
at
generic points in the moduli space. This moduli space contains enhanced
symmetry points with higher level realizations of the gauge symmetry:
$SU(9)_2$,
$(SU(5)_2)^2$, and $(SU(3)_2)^4$. Twisting the $(SU(5)_2)^2$ solution gives a
solution with the orthogonal gauge group $(SO(9)_2)^2$.

We have constructed fermionic realizations of a large range of
four-dimensional N=4 and six-dimensional N=2 supersymmetric solutions to the
heterotic string with
semi-simple groups of varying rank, containing both simply-laced and
non-simply-laced factors,
and with part or all of the gauge symmetry realized at higher level.
It should be stressed that four-dimensional N=4 supersymmetry need not always
arise
via toroidal compactification from a higher dimensional theory.
The clearest evidence for this is the existence of an N=4 four-dimensional
solution where the gauge symmetry is reduced to the minimum consistent with the
world-sheet supersymmetry constraints. Its fermionic realization uses a
spin structure block of 44 left-moving real fermions. The
number of abelian vector multiplets in this N=4 theory is just {\it six}.

The development of fermionization techniques \cite{klst}\cite{cchl} has
enabled the efficient sampling of new classes of exact solutions to string
theory.
It is important to focus on those aspects of the solutions that have generic
implications for our understanding of string theory. We would like to stress
that
there exist additional maximally supersymmetric solutions, in any space-time
dimension, which do {\it not} have fermionic realizations. A simple example in
D=4 is toroidal compactification with gauge group
$(SU(3))^3$$\times$$E_8$$\times$$E_8$. The particular choices of affine Lie
group, rank,
or Kac-Moody level, obtained in the fermionic construction should not be
emphasized.
On the other hand the existence of maximally supersymmetric theories with
distinct
target space duality groups, and the fact that non-simply-laced and
simply-laced
gauge groups enter on an equal footing, are generic observations
relevant for further study.

In concluding, we note that the construction of alternative
four-dimensional N=4 string theories is a useful step towards
constructing simpler
pedagogical models for studying the low energy physics of string theory.
Any four-dimensional N=1 supersymmetric solution to heterotic string theory
inherits some of its structure from a parent N=4 solution.
Four dimensional N=4 heterotic string theories of lower rank are a
more appealing starting point for constructing pedagogical N=1 models
via twisting, because the models will inherit a
reduced massless particle spectrum. In toroidal
compactifications the massive modes of the string spectrum were completely
determined by the low energy symmetries: extended supergravity and
gauge symmetry. This is no longer true in the maximally supersymmetric
solutions
constructed in this paper. The mechanism by which finiteness is achieved in
these solutions does not rest wholly upon the finiteness of the low-energy
field theory limit. Twisting such solutions to construct pedagogical N=1 models
with chiral matter will teach us about new, and intrinsically stringy,
mechanisms
for achieving finiteness.

We acknowledge helpful discussions with Jeff Harvey, Joe Polchinski,
Ashoke Sen, and Andy Strominger. S.C. thanks Keith Dienes for
verifying the absence of tachyons, spacetime supersymmetry,
and degeneracies of physical states at both the massless and massive
levels in the four-dimensional N=4 $Sp(20)$$\times$$SO(9)$ solution.
J.L. thanks Stephen Chung for discussions.
Additional details of the solutions described here can be accessed
at http://www-theory.fnal.gov/superstrings/superstrings.html.
This work was supported in
part by NSF grant PHY91-16964 and by DOE grant DE-AC02-76CH03000.


\begin{thebibliography}{99}

\bibitem{seiberg} S. Mandelstam, {\it Nucl.\ Phys.}\/
{\bf B213} (1983) 149; L. Brink, O. Lindgren, and B. Nilsson,
{\it Nucl.\ Phys.}\/ {\bf B212} (1983) 401; N. Seiberg,
{\it Phys.\ Lett.}\/ {\bf 206B} (1988) 75.

\bibitem{gno} C. Montonen and D. Olive, {\it Phys.\ Lett.}\/ {\bf 72B} (1977)
117;
E. Witten and D. Olive, {\it Phys.\ Lett.}\/ {\bf 78B} (1978) 97;
H. Osborne, {\it Phys.\ Lett.}\/ {\bf 83B} (1979) 321.

\bibitem{gira} J. Harvey, G. Moore, and A. Strominger,
{\it Reducing S-duality to T-duality}, hep-th/9501022;
L. Girardello, A. Giveon, M. Porrati, and A. Zaffaroni,
{\it S-duality in N=4 Yang-Mills Theories with General Gauge Groups},
preprint NYU-TH-94/12/1, hep-th/9502057.

\bibitem{sei} N. Seiberg, {\it Electric-magnetic Duality in Supersymmetric
Non-abelian Gauge Theories}, preprint RU-94-82, hep-th/9411149; K.
Intriligator
and N. Seiberg, preprint RU-95-3, hep-th/9503179.

\bibitem{sen} A. Sen, {\it Int.\ Jour.\ Mod.\ Phys.}\/ {\bf A9} (1994) 3707;
J. Schwarz and A. Sen, {\it Nucl.\ Phys}\/ {\bf B411} (1994) 35.

\bibitem{town} C. Hull and P. Townsend, {\it Unity of string dualities},
preprint QMW-94-30, hep-th/9410167.

\bibitem{witten} E. Witten, {\it String Theory Dynamics in
Various Dimensions}, preprint IASSNS-HEP-95-18, hep-th/9503124.

\bibitem{harstrom} A. Sen, {\it String-string duality conjecture in
six dimensions and charged solitonic strings}, preprint TIFR-TH-95-16,
hep-th/9504027; J. Harvey and A. Strominger, preprint EFI-95-16,
hep-th/9504047.

\bibitem{narain} K. S. Narain, {\it Phys.\ Lett.}\/ {\bf 169} (1986) 41.
K. S. Narain, M. Sarmadi and E. Witten, {\it Nucl.\ Phys.}\/ {\bf 279B} (1987)
369.

\bibitem{ginsparg} P. Ginsparg, {\it Phys.\ Rev.}\/ {\bf D35} (1987) 648.

\bibitem{bd} T. Banks and L. Dixon, {\it Nucl.\ Phys.}\/ {\bf B307} (1988) 93.

\bibitem{klst} H. Kawai, D. Lewellen, J. Schwartz, and S.-H. H. Tye,
{\it Nucl. Phys.}\/ {\bf B299} (1988) 431.

\bibitem{cchl} S. Chaudhuri, S.-w. Chung, G. Hockney and J. Lykken, {\it
String Consistency for Unified Model Building}, preprint Fermi-pub-94/413-T,
hep-ph/9501361.

\bibitem{salam} A. Salam and J. Strathdee, {\it Ann. Phys.} {\bf 141} (1982)
316.

\bibitem{agwit} L. Alvarez-Gaume and E. Witten, {\it Nucl. Phys.} {\bf B234}
 (1983) 269.

\end{thebibliography}
\end{document}